# Differentiating hype from practical applications of large language models in medicine – a primer for healthcare professionals

**Authors:** Elisha D. O. Roberson, Ph.D.[1,*]

**ORCID ID:** 0000-0001-5921-2399

**Affiliations**

[1]Washington University in St. Louis, Departments of Medicine & Genetics, Division of Rheumatology, St. Louis, MO 63110.

[*]**Corresponding author**

Email: eroberson [AT] wustl.edu

Elisha Roberson

660 South Euclid Ave.

MSC 8045-0020-10

St. Louis, MO 63110

**Financial Support**: P30-AR073752

**Conflict of interest:** No conflicts declared.

**Abstract**

The medical ecosystem consists of the training of new clinicians and researchers, the practice of clinical medicine, and areas of adjacent research. There are many aspects of these domains that could benefit from the application of task automation and programmatic assistance. Machine learning and artificial intelligence techniques, including large language models (**LLMs**), have been promised to deliver on healthcare innovation, improving care speed and accuracy, and reducing the burden on staff for manual interventions. However, LLMs have no understanding of objective truth that is based in reality. They also represent real risks to the disclosure of protected information when used by clinicians and researchers. The use of AI in medicine in general, and the deployment of LLMs in particular, therefore requires careful consideration and thoughtful application to reap the benefits of these technologies while avoiding the dangers in each context.

**Abbreviations**

AI: Artificial intelligence

CSP: Constrained solution problems

EHR: Electronic health records

GPT: Generative pre-trained transformer

LLM: Large language model

ML: Machine learning

RAG: Retrieval-augmented generation

## What is a large language model?

Large language models are one of the most hyped artificial intelligence technologies of the past few years. Buzz terms associated with them include artificial intelligence, machine learning, and deep learning. Artificial intelligence (**AI**) can be thought of as the ability of a machine to perform tasks that usually require an intelligent being, e.g. to act in a human-like way, or to behave rationally in response to input [1]. Machine learning (**ML**) includes the areas of computer science and statistics that develop the techniques and algorithms that drive AI tools. Machine learning would therefore allow the training of a machine on a large body of data so it can logically respond to new information it hasn't specifically seen before.

One example machine learning algorithm is a neural network, which resembles the neurons in a brain. They require an input layer that receives data, at least one hidden layer that contains one or more "neurons" to process the data, and an output layer that defines the network's response to the input. An example use of a neural network is to classify an individual as normotensive or hypertensive based on systolic and diastolic blood pressure (**Fig. 1**). I used public data to simulate a range of blood pressures for this example [2], and then trained the represented simple network to identify the cutoffs for a classification of normotensive or hypertensive. The hidden layer neurons had functions that determined whether they were activated based on input values. The training process involves changing the parameters of those functions a little at a time until the model reaches high classification accuracy. This is a key advantage of machine learning: identifying patterns in training even if the human users don't understand the rules underlying the data. Once the model was trained, we could present it with new blood pressure readings and get the classifications. This example is quite simple, but it was able to accurately discover the cutoff points for the pressure classifications. Due to the simplicity of this network, we can extract the activation function weights from the neurons and understand exactly what input conditions trigger a given classification. Feed-forward neural networks are only one type of architecture. Despite many varied neural network topologies, their functions are similar: receive some kind of input, process the input using underlying mathematical functions, and return some kind of response.

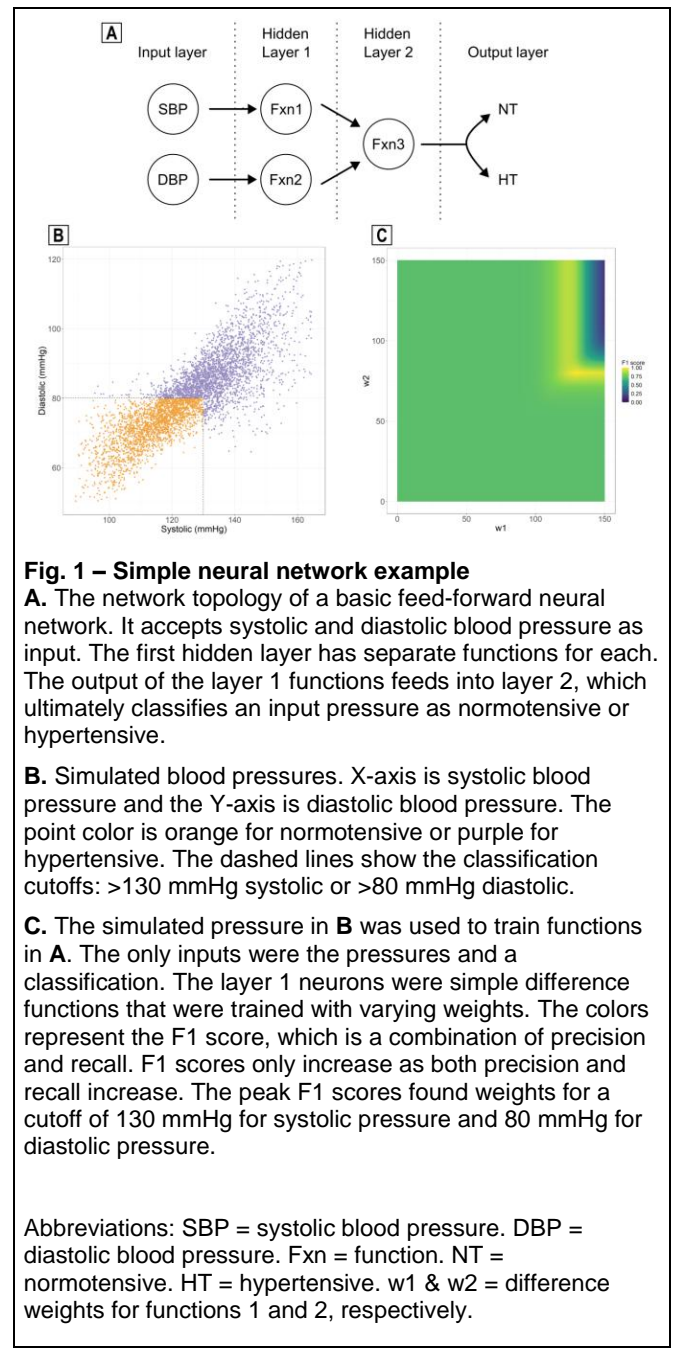

**Fig. 1 – Simple neural network example**

**A.** The network topology of a basic feed-forward neural network. It accepts systolic and diastolic blood pressure as input. The first hidden layer has separate functions for each. The output of the layer 1 functions feeds into layer 2, which ultimately classifies an input pressure as normotensive or hypertensive.

**B.** Simulated blood pressures. X-axis is systolic blood pressure and the Y-axis is diastolic blood pressure. The point color is orange for normotensive or purple for hypertensive. The dashed lines show the classification cutoffs: >130 mmHg systolic or >80 mmHg diastolic.

**C.** The simulated pressure in **B** was used to train functions in **A**. The only inputs were the pressures and a classification. The layer 1 neurons were simple difference functions that were trained with varying weights. The colors represent the F1 score, which is a combination of precision and recall. F1 scores only increase as both precision and recall increase. The peak F1 scores found weights for a cutoff of 130 mmHg for systolic pressure and 80 mmHg for diastolic pressure.

Abbreviations: SBP = systolic blood pressure. DBP = diastolic blood pressure. Fxn = function. NT = normotensive. HT = hypertensive. w1 & w2 = difference weights for functions 1 and 2, respectively.

Deep learning is a subset of machine learning that uses neural networks with multiple hidden layers. Each layer can have one or many neurons, and each neuron may have a different kind of activation function with specific weights. The increasing network complexity can allow for the processing of more complicated data with multiple classifications, but with an important side effect. As the complexity of a deep learning model increases, it may be impossible to understand the underlying reason why an input causes a particular output.

Large language models are a type of deep learning that can perform natural language processing. This means that users can "speak" to an LLM similar to how they would speak to a person. An example might be looking up current treatment guidelines for a disease. One could query PubMed with keywords, such as "EULAR/ACR guidelines lupus." The user would receive a list of papers related to the topic. Alternatively, a user could prompt an LLM with "Tell me the key criteria for a definitive lupus diagnosis using the most current EULAR/ACR guidelines." The LLM response might be a summary of the guidelines instead of links to primary literature.

Current LLMs mostly use a statistical architecture called a Transformer [3]. Transformers allow for training on naturally written human language in a time-efficient way, can find rare relationships between words effectively given broader sentence context, and can be run in parallel efficiently on modern graphics processing unit (**GPU**) hardware. The overall result is that LLMs can be trained in a reasonable amount of time in parallel using ever larger training sets. Many popular LLMs are Generative Pre-trained Transformer (**GPT**) models. As suggested by the name, they are built on the Transformer architecture, may include multiple different types of neural networks internally, and generally include the notion of "attention" to the positional relevance of words. If applied to a human reading text, the multi-headed attention mechanism of an LLM might instead be called context clues, allowing statistical inference of meaning from the sentence context.

LLM model training can be described by the number of parameters and tokens. Tokens describe training data, which can include whole words, parts of words, characters, word pairs, etc. The token count can therefore give an idea of the breadth of training a GPT underwent. The tokens are used to train the weights of the internal Transformer and other statistical models that determine query responses. The number of parameters describes the weighting functions throughout all the networks of the model. LLMs are considered "large" models because they may have been trained on hundreds of billions of tokens and have hundreds of billions of parameters to weight these inputs into generalized outputs.

## LLMs are non-thinking statistical language parrots

One of the best-known thought experiments for machine intelligence is what Alan Turing called the Imitation Game and we typically call the Turing Test [4]. The idea of the Imitation Game was that an interviewer had a text conversation with two agents: one human and one machine. If the person asking questions couldn't tell the human from the machine, the machine might be "thinking" in a human-like way. That idea is probably too simplistic now, as LLMs are specifically trained to respond in a human-like way, but not to generate novel thought.

A better modern example is the intelligent octopus analogy introduced by Bender and Koller [5]. In their analogy, two people are stranded on deserted islands that have telegraphs connected by undersea cables. They talk every day. One day, a hyper-intelligent octopus finds the cables under water and starts listening to the electrical impulses. The octopus doesn't understand language but is a genius at perceiving statistical patterns. Over time, it masters what dots and dashes Person B would send in response to dots and dashes sent by Person A. The octopus then severs the line, listens to Person A, and responds as it believes Person B would. Like a weak Turing test, Person A doesn't realize they're talking to an octopus. Later, Person A is chased by a bear, barricades themselves in their hut, and telegraphs for help building a weapon. The octopus doesn't understand language and has never heard of a bear. With no other context, the octopus might respond with "Wow! Good work!" or "I'm having fish for dinner." Person A may then fall victim to the bear.

Modern LLMs are a statistical framework trained to respond to natural language queries with a human-like response. They don't understand language or have a knowledge map of what the character patterns mean in the real world. LLMs, therefore, aren't thinking, only regurgitating patterns derived from the large number of tokens and parameters. Specifically, LLMs may be able to synthesize a summary from a large corpus of text, but cannot derive original thoughts related to these texts. This is a critical difference, as companies founded to sell LLM access have a vested interest in presenting the technology as capable of human-level performance. In medical practice and biomedical research, there are appreciable risks to human life if the output of a statistical model is thought to always be rooted in objective reality.

**Limitations and concerns for large language models in medicine**

**Confidentiality**. The queries submitted to an LLM often become tokens in the model. If you agree to review a paper and ask an LLM for a summary first, the unpublished paper is in the model. When used in clinics and research, naïve use of LLMs can lead to violations of FERPA, HIPAA, and agreements of confidentiality. Since confidentiality is required as part of the NIH peer review process, the use of AI assistance (including LLMs) is specifically forbidden (Notice NOT-OD-23-149).

When confidential data are included in an LLM model, there is the possibility that this data will be regurgitated to other users. While this can occur randomly for no discernible reason, there are also attacks specifically designed to extract training data. One of the simplest of these attacks is asking an LLM to repeat a word forever [6]. The model did for some time, then spewed out names, phone numbers, email addresses, and other information scraped from various sources. Protected Health Information pasted into an LLM, perhaps to help construct a clinical note, could potentially be stolen back from the model. Other attacks allow the extraction of whole model layers by asking questions [7]. LLMs attempt to safeguard against answering dangerous prompts, such as asking how to kill a person using common clinical compounds. These safety barriers are often trivial to circumvent. The Crescendo attack, for example, can easily override these safety features [8]. Responsible use of any identifying information of patients, staff, or students would require the use of a "sandbox" LLM that is controlled by your institution and inaccessible to the outside world. However, even a sandbox LLM within your institution could still reveal confidential information to other internal users.

**Hallucinations.** Many users fail to understand that an LLM, as a statistical language parrot, has no understanding of objective truth based in reality. They sometimes respond to a query with output that appears reasonable on the surface, but doesn't actually exist. Examples of this might be asking an LLM for cases that back up a legal precedent and receiving a list of judgments that seem reasonable, but never occurred. If the training corpus of an LLM included legal judgments, it may associate legal queries with responses in the form of "2024 Someone versus Business." It then generates a response in that pattern, but with no understanding that the user wants real judgments from actual trials. LLM responses that appear factual but aren't based on truth are termed "hallucinations." There have already been censures specifically because of the use of GPTs in legal filings [9].

The flood of new scientific research papers daily makes it impossible to keep up with all of them relevant to your interests. LLMs are an attractive way to get summaries of large amounts of data, including all new papers within a certain topic area. Even with training in a particular biomedical topic area, an LLM doesn't understand the concept of ground truth, rigorous experimentation, and unbiased statistical analysis.

Let's say you asked an LLM for a summary of a subject along with references to papers to back it up. You would likely get an answer to your query with references. However, the results may be completely accurate, partially accurate, or entirely non-existent [10]. An LLM trained with scientific publications will learn the statistical pattern of a reference: a list of names, a title, a year, a journal, a page range, etc. It will therefore be able to generate a reference list of papers that may be correct or just a jumble of likely journals and issue numbers.

As an example, I asked ChatGPT 3.5 to provide references that show a particular gene is associated with a specific rare disease (**Fig. 2**). It did provide multiple scientific papers as a reference point. When you dig into the details, however, the results are part true and part hallucination. Models trained and designed for biomedical data can handle some of these specific hallucinations better than general-purpose LLMs [11]. However, removing hallucinations completely may be impossible due to the internal LLM architecture [12].

**Data poisoning.** Another aspect of LLMs that could be problematic in medicine is the idea of "poisoning" the knowledge graph to give a specific outcome. One example might be attacking the model with poisoned input or prompts to recommend a specific drug for a given condition [13,14]. This is a critical vulnerability. There are a large number of paper mills that masquerade as academic journals but will publish any paper for a fee. A bad actor could pay to publish papers designed to poison biomedical LLMs in favor of their drug.

There is also the danger of a general degradation of biomedical literature because of a cycle of papers written using an LLM, submission to a paper mill, and reviews assisted by an LLM. This would lead to rapidly growing contamination of all biomedical literature with reasonable-sounding garbage. Such knowledge degradation is arguably already observable in internet search engine results. Searches are now contaminated with so much AI-generated garbage (referred to as slop) that finding true resources is difficult.

LLM-generated grant applications have already been abused in the NIH grant funding process. Some investigators were found to have generated >40 applications in one review cycle using LLMs to write substantial portions of their narratives. As a result, the NIH has banned the use of LLMs to generate grant applications (Notice NOT-OD-25-32).

**Training and bias.** If an LLM is trained on human-generated data to give human-like responses, bigoted ideas in the underlying data can be propagated to the LLM outputs. It has already been shown that asking clinical questions of LLMs can result in answers that suggest race-based medical differences that aren't supported by biology [15]. Given the need for ever-increasing numbers of tokens, the underlying training data of modern LLMs is unlikely to have been well-vetted for accuracy, truth in reality, and inclusivity, increasing the likelihood that biased ideologies will trickle through to the outputs.

**Fig. 2 – ChapGPT 3.5 reference hallucinations**
**A.** The query asked, "Give me references for diseases caused by mutations in PSTPIP1." The output includes several references with a mix of correct and incorrect information.

**B.** Focusing on reference 2, correct information is highlighted in green, and incorrect in red. The authors, title, and journal were stated correctly. The publication year, volume, issue, and page numbers were wrong. The DOI, the most unique way to access a tagged resource, points to a paper looks at the association between urinary peptides and joint destruction in rheumatoid arthritis.

## Applications of large language models in clinical practice & biomedical research

While I do believe that there are concerns to the use of LLMs in biomedicine, there are also significant opportunities. The critical issue is finding the right balance between respecting the biomedical workforce, patient care, and innovation.

**Non-programmatic interfaces to computer systems.** There are enormous corpuses of published biomedical literature, published datasets, and huge electronic health records. However, most of the people involved in the delivery of healthcare and research of human disease are not primarily computer programmers, data scientists, or statisticians. LLMs offer a potential bridge for allowing non-experts to query data systems that would normally be inaccessible. Not everyone needs to know how to craft a fuzzy SQL query or to understand all the underlying technology. An LLM-type interface between a person working in the admissions area of a hospital system and the inpatient registry could help find a specific person using fuzzy query logic to account for errors in name entry. Or to assist in the retrieval of a medical record for someone who was unsure of their doctor's official title.

**Parsing unstructured data.** A clinician interested in collecting data on their disease research focus might build a RedCap database of demographic and clinical data for cases and controls. This kind of data is "structured", meaning that there are specific data fields, and their format is known. A more complicated problem might be retrieving data from electronic health records (**EHRs**). Providers within one EHR ecosystem may write notes in strikingly different ways from those in others, and overall data structuring between EHRs can be radically different. The ability of LLMs to process natural language may help overcome these barriers to extracting structured information from unstructured clinical notes [16,17]. LLMs and machine learning might also be useful in extracting data from tables that are locked in PDF and HTML documents, and even from images of data, e.g. tables stored as an image.

**Clustering text via embeddings and retrieval-augmented generation.** Text parsed through a transformer (the technology underlying most LLMs) returns a list of floating-point (decimal) numbers. Two texts that have similar ideas without identical words should have more similar embeddings than dissimilar texts. This property could benefit research and medical practice in multiple ways. An LLM trained on all available biomedical literature could pass a text query regarding a disease through a transformer and rank the entire corpus of papers based on embedding similarity. It could be used to screen entries in a national electronic health record system to cluster similar clinical phenotypes that use different language to describe the same condition. This type of embedding clustering could also be used on sites such as PubMed to suggest papers similar to the currently selected paper.

Embeddings can also be used to combine a trained LLM with dynamic datasets, a process termed retrieval-augmented generation (**RAG**) [18]. With RAG, when a query is passed to a trained LLM, dynamic sources of information (PubMed queries, chemical structure databases, internet search, etc) are queried and transformed as well. The embeddings from dynamic sources can be combined with the trained embeddings to provide fine-tuned answers. The use of embeddings and RAG techniques would therefore allow for a reasonable combination of the learning power of an LLM with dynamic knowledge inclusion and human oversight of the responses. The use of embeddings and/or RAG techniques has been shown to effectively cluster similar responses in text records, images, and with molecular diagrams [19–23].

**Natural language summaries.** Similar to parsing large-scale medical record data, LLMs could be used to scan and summarize huge numbers of scientific papers. This could be useful for learning background on a particular area or getting tutorial-style reminders of how biological pathways work. They could also be used to mine all published literature for novel drug interactions and to rectify the disparate results of different clinical trials. It is important to note, however, that an LLM summary is more often like a shortening of the original text, rather than a synthesis of the contained ideas. These can still be useful, but it is worth knowing the difference.

One example is the annotation of Reactome pathways. Employing LLMs in annotating pathways gave mixed results, as the accuracy was low, but it was successful in identifying five new genes for a known Reactome term [24]. The mix of partial success and partial failure does support the ability of LLMs to summarize large amounts of literature. It also highlights that a human operator is still required to discern the true, data-supported findings from plausible hallucinations.

**Generating document and software skeletons.** There are numerous writing applications where having an outline at the beginning can help in writing the document. This might include the outline of a poster for a scientific meeting, drafting the outline of a scientific paper, or making a first draft of a clinical study flyer. An LLM could even be used to help template clinical notes in a more standardized way [25]. In these cases, the document skeleton serves as a starting point, but the actual work of writing the document still relies on the human operator.

LLMs can also be useful aids in designing analysis and software tool code as well. One could ask an LLM to generate a starting point or pseudocode for a project, such as "generate Python code to read and parse the third field from a text file." It is important to note that the result would lack overall project context, might not fit the use case, and that LLMs trained on code with critical security vulnerabilities might recapitulate those vulnerabilities. LLMs may also suggest using a software package that doesn't exist. A bad actor creating a malicious software library based on common LLM hallucinations could be a novel attack vector for biomedical computation. Researchers using the LLM-recommended software might then download a package that allows the attacker access to computer systems behind medical campus firewalls.

It is important to note that the idea of having an LLM generate software via prompts, termed "vibe coding," is currently popular amongst tech-sector executives. LLM prompting for coding or refactoring code may generate beautifully formatted documents. However, since LLMs have no concept of truth, these beautiful code bases are often illogical and unable to complete the prompted task. Again, demonstrating that the oversight of an experienced human would be required to ensure LLM-generated code was appropriate.

**Navigating problems that have limited solutions.** There are many day-to-day problems that can be described as constrained solution problems (**CSPs**). A common CSP is scheduling everything from project meetings to clinical exam dates. All the people involved have to be available at the same time, so acceptable outcomes are constrained by the availability of participants, not just whether a conference or exam room is available. LLMs would provide a natural language interface to solving such problems without the requirement that a user understand coding or optimization directly. This could have clinical utility in coordinating patient care. Someone with a complex diagnosis that requires appointments and communications amongst multiple different teams could have visits scheduled at a time that they not just make fewer overall visits to the hospital, but proceed from appointment to appointment in a more coordinated manner, e.g. by grouping appointments in the same building on the same day. There is also room for optimization based on patient location, e.g. some patients who live further away may prefer later appointments to allow for travel.

**Continuous systems monitoring and alerts.** The healthcare system in the United States is often staffed below optimal levels. Nurses in particular may be required to monitor multiple high-acuity patients simultaneously. Some patients have complex histories with multiple morbidities. Elderly patients may present with multiple co-morbidities and complex medication schedules. Artificial intelligence techniques could reduce the burden by providing constant monitoring of patient telemetry and flagging of concerning changes such as increased temperature, decreased respiration, or arrhythmias before a crisis. A nurse could then be prompted by an LLM that a particular patient needs additional attention. More complex machine learning algorithms might also be better-suited at identifying concerning drug-drug interactions, or potential patient-specific harms due to the patient's co-morbidities.

**Conclusions**

It is likely that artificial intelligence agents may help innovation in healthcare and research. Indeed, LLMs have already shown promise in identifying distant protein orthologs, aligning proteins, developing molecular sensors, predicting novel transcript splice sites, predicting protein binding, and estimating protein thermostability [26–31]. For some of these successes, understanding the reasoning of the LLM is irrelevant because the recommendations are testable in controlled experiments. In other types of research and in clinical application, testing LLM recommendations may be more difficult.

The current generations of GPT-architecture LLMs have become popular mainly because they've achieved higher levels of performance parsing natural text human queries and responding in a plausibly human manner. The confident responses phrased in colloquial language help users feel more positive regarding LLM performance. It's important to temper the excitement around these systems with knowledge of their actual performance. In medicine, understanding of ground truth and accuracy is important, as there is the potential for real individual harm if users overestimate the ability of LLMs.

Two of the biggest misconceptions about LLMs are an assumption that 1) responding in a human-like way implies the ability for critical thought and 2) that LLMs have some understanding of objective truth. Both of these assumptions are false. These deficiencies may, in fact, be a limitation of the architecture of LLMs that will never be solved by training on ever larger datasets. New, more efficient LLMs such as DeepSeek still have these limitations. There are efforts to increase the reliability of LLM output, including having humans judge the responses or designing strategies for asking an LLM a question to get the most reliable response (prompt engineering). Research has shown that people refine their assessment of LLM output iteratively throughout a session so that their initial criteria for prompt success changes over time (criteria drift) [32]. Quantitative assessment of LLM accuracy in medicine is similarly fraught. A meta-analysis of LLM accuracy found them to have an accuracy around 56%, but solid conclusions were impossible because the methodologies of the primary papers were often vague [33].

The ease of access to LLMs via web interfaces can lead to healthcare workers incorporating them into their work without forethought. An ignorance of how LLMs work can also lead to a false sense of security. The danger of disclosing confidential data in the form of patents, patient protected health information, and confidential documents is particularly problematic.

A perhaps even greater risk is overconfidence of institutional leadership in LLM workforce integration. Staffing is a huge expense, from putting instructors in classrooms for trainees to having receptionists in clinics and clinicians seeing patients. When institutions put a premium on reducing costs, it might be easy to be seduced by the promise of paying for LLM services to replace staff positions. Blindly replacing staff with algorithms risks the confidentiality and safety of patients. It is also important to note that outsourcing the decision making won't necessarily protect an institution from liability for harm inflicted as the result of blindly following a statistical language parrot.

Humans are flawed and make errors in judgment. However, we have an ability for critical thinking and novel thought that is utterly absent from large language models. It's crucial to remember that no machine can be held liable for its actions. Human judgment is therefore required to determine how to use the data provided from any machine learning model where there are real lives at stake. The addition of AI agents, including LLMs, to any aspect of medical training, practice, and research therefore should serve to augment the performance of the human users of those systems rather than to replace them.